\documentclass[%
reprint,
amsmath,amssymb,
aip,
]{revtex4-2}

\usepackage{graphicx}
\usepackage{dcolumn}
\usepackage{bm}
\usepackage{xcolor}

\newcommand{\changes}[1]{\textcolor{black}{#1}}

\newcommand{\ve}[1]{\mathbf{#1}}
\newcommand{\bvec}{\begin{pmatrix}}
	\newcommand{\evec}{\end{pmatrix}}

\graphicspath{{figs}}

\begin{document}
	
	
\title{Enhancement of Ponderomotive End Plugs with Low-Mass Dopants}

\author{Ian E. Ochs}
\email{iochs@princeton.edu}
\author{Elijah J. Kolmes}
\author{Nathaniel J. Fisch}%
\affiliation{Department of Astrophysical Sciences, Princeton University, Princeton, NJ 08544}%

\date{\today}

\begin{abstract}
	During conventional confinement of a linear plasma using ponderomotive end plugs, a repulsive potential is created at the axial ends by employing a perpendicularly-polarized wave with a wave frequency greater than the ion gyrofrequency.
	This potential is then partially cancelled out by an ambipolar potential that arises to equilibrate electron and ion densities along the field line.
	However, recent work on centrifugally-confined plasmas has shown that the appropriate choice of a dopant in the end region can dramatically change the structure of the ambipolar potential.
	For the ponderomotive potential, this ambipolar shaping can be even more powerful, since a lightweight dopant can see a different sign of ponderomotive potential relative to the confined species. 
	As a result, the dopant can reverse the sign of the ambipolar potential, causing it to  dramatically enhance the end plug ponderomotive confinement.
\end{abstract}

\maketitle


\section{Introduction}

In a linear plasma, such as a mirror machine, applying a confining potential to the ion population can dramatically improve the confinement.
For instance, rotating the plasma can lead to a centrifugal potential that pushes the ions toward the midplane.\cite{Lehnert1971RotatingPlasmas,Bekhtenev1980ProblemsThermonuclear,Ellis2005SteadySupersonically,Teodorescu2010ConfinementPlasma,Kolmes2024MassiveLonglived}
Similarly, applying radio-frequency (RF) electromagnetic fields at the ends of the device can result in ponderomotive potentials that repel the ion species.\cite{Gaponov1958PotentialWells,Pitaevskii1961ElectricForces,Motz1967RadioFrequencyConfinement,Hidekuma1974PreferentialRadiofrequency,Hiroe1975RadiofrequencyPreferential,Watari1978RadiofrequencyPlugging,Weibel1980SeparationIsotopes,Dimonte1982PonderomotivePseudopotential, Anderegg1995LongIon,Dodin2005ApproximateIntegrals,Miller2023RFPlugging,Scheffel2025AxialConfinement}
Such forces can also be generated in a rotating plasma by static electrodes, which appear as oscillating waves in the moving frame.\cite{Rubin2023MagnetostaticPonderomotive,Ochs2023CriticalRole,Rubin2024FlowingPlasma,Rubin2025PonderomotiveBarriers}

Because electrons are low-mass, they generally do not see these ion-focused centrifugal or ponderomotive forces.
Thus, in order to enforce quasineutrality along a magnetic field line, electric potentials develop along the field line to eliminate the pileup of space charge.
In a single-species plasma, these ``ambipolar'' potentials typically have the effect of partially screening the confining potential.

However, it was recently shown\cite{Kolmes2025IonMix} that in a plasma with multiple ion species present, the effects of the ambipolar potentials can be far more dramatic.
In particular, the ambipolar potentials can grow large enough to overwhelm the centrifugal potentials, causing certain ion species to be expelled from regions of low centrifugal potential.
This reversal of the centrifugal force allows for new types of ``doped'' centrifugal and ambipolar end cells, which manipulate the ambipolar potential by introducing ions of different masses and charge states.

In this paper, we examine similar ambipolar manipulation using the RF-induced ponderomotive potential to produce enhanced ponderomotive end plugs.
The main extension over the centrifugal case is that the ponderomotive potential, unlike the centrifugal potential, can take a different sign for different ion species.
This sign switch allows for even greater end-plug enhancement, as it allows the ponderomotive and ambipolar potentials to stack.

To understand the basic mechanism for the enhancement, recall that the ponderomotive potential $\Psi$ experienced by an ion interacting can take  a different sign depending on the wave frequency $\omega$ and cyclotron frequency $\Omega \equiv Z e B/ m c$. In particular, for a linear wave polarization perpendicular to the magnetic field $\ve{B}$:
\begin{align}
	\Psi \propto \frac{Z^2}{m} \frac{1}{1 - \bar{\Omega}^2}; \quad \bar{\Omega} \equiv \frac{\Omega}{\omega} \propto \frac{Z}{m}, \label{eq:PonderomotiveScaling}
\end{align}
where $Z$ is the ion charge state, and $m$ is the mass.
\changes{[Note: this a special case of a more general potential based on the circular polarizations, as discussed in Appendix~\ref{app:GeneralPolarization}].}
In conventional ponderomotive end-plugging, we can then produce confinement for a confined ion species $c$ by localizing a wave with $\bar{\Omega}_c < 1$ in the end cell of the plasma.
Usually, the ambipolar field will then adapt to equate electron and ion densities in the end cell, partly screening the ponderomotive potential.

\begin{figure}
	\centering
	\includegraphics[width=\linewidth]{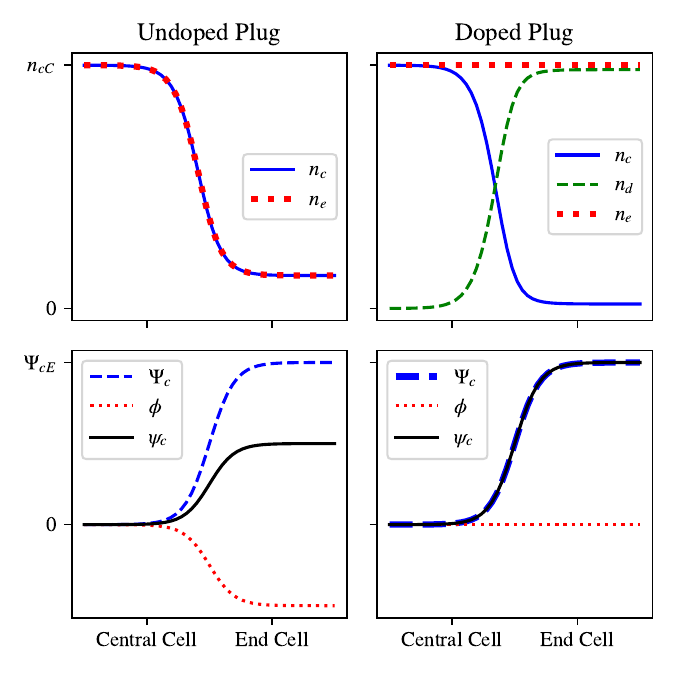}
	\caption{Schematic plots showing the principle behind a doped ponderomotive end plug.
	In an undoped ponderomotive end plug (left column), an RF wave above the confined species cyclotron frequency $\Omega_c$ is applied in the end cell region, resulting in a repelling potential $\Psi_{c}$ (bottom left, dashed blue line) maximized at $\Psi_{cE}$ in the end cell . 
	However, because the electron density $n_e$ (top left, red dotted line) must match the ion density $n_c$ (top left, blue line) to avoid large space charge effects, an ambipolar potential $\phi$ (bottom left, red dotted line) forms which partially cancels the ponderomotive potential.
	This results in a lower net confining potential $\psi_c$ on the confined ion species.
	In contrast, for a doped ponderomotive end plug (right column), an additional ion species (top right, green dashed line) is added, which is attracted to the end cell by the RF wave.
	This flattens the electron density (top right, red dotted line), zeroing out the ambipolar potential (bottom right, red dotted line).
	As a result, the net confining potential $\psi_c$ is no longer reduced.
	If more dopant is added, the electron density can actually be higher in the end cell than in the central cell, reversing the ambipolar potential relative to the undoped case and effectively acting as a tandem cell.}
	\label{fig:DopingSchem}
\end{figure}

However, if we dope the end cells of the plasma with a different ``dopant'' species $d$ for which $\bar{\Omega}_d >1$, then species $d$ will be attracted to the end cell.
The presence of this ion dopant will cause the ambipolar potential to rise in the end cell as electrons try to co-localize with the dopant (Fig.~\ref{fig:DopingSchem}).
This ambipolar rearrangement has the potential to reduce the electron screening substantially, or even to enhance the repulsive potential experienced by the confined species $c$ over the unscreened ponderomotive potential.
In the latter case, the doped end plug acts effectively as a tandem cell,\cite{Dimov1976ThermonuclearConfinement,Fowler1977TandemMirror,Post1987MagneticMirror,Fowler2017NewSimpler} with the electrons held in place by the charge of confined doped ions.
\changes{Importantly, this approach can lead to substantial enhancement of the ponderomotive confinement even when the dopant density in the central cell is negligible, which means that confinement might be achieved in a fusion reactor without ``poisoning'' the fusion reaction.}

The paper is organized as follows.
In Sec.~\ref{sec:SingleSpecies}, we review the quasineutrality condition, and show how it gives rise to ambipolar screening of the ponderomotive potential in a single-species plasma.
In Sec.~\ref{sec:HighDoping}, we make further use of this single-species theory, showing how in the limit of high dopant density, the ponderomotive potential can be substantially enhanced by localizing a lightweight ion dopant in the end cell.
In Sec.~\ref{sec:LowDoping}, we examine the opposite limit of low dopant density, allowing us to establish the minimum dopant density required to achieve substantial confinement enhancement.
Finally, in Sec.~\ref{sec:Numerics}, we validate the analytic theory via full numerical solution of the quasineutrality equations, showing that substantial confinement enhancement is possible even with negligible dopant density in the central cell.

\section{Ambipolar Screening} \label{sec:SingleSpecies}

As it travels along a magnetic field line, a particle of species $s$ sees a total potential that consists of both electrical and ponderomotive terms:
\begin{align}
	\psi_s = \Psi_s + Z_s e \phi. \label{eq:psiS}
\end{align}
Generally, $\Psi_s$ is externally applied, while $\phi$ is self-consistently determined to enforce ambipolarity.

To simply determine $\phi$, we take all species to be Maxwellian, ignoring the effects of both nonuniform magnetic fields and the loss cone.
We also take the plasma midplane in the central cell to represent the 0-point of all potentials in the plasma.
As a result, the density of species $s$ anywhere in the plasma is given by:
\begin{align}
	n_{s} = n_{sC} e^{-\psi_s/T_s},
\end{align}
where $T_s$ is the temperature of species $s$, and $n_{sC}$ is its density at the central cell midplane.
Then, \changes{because the plasma constituents can move freely to shield any electric fields produced by charge imbalances}, the total charge density at any point of the plasma must be approximately zero, i.e.
\begin{align}
	0 &= \sum_s Z_s n_s = \sum_s Z_s n_{sC} e^{-\psi_s/T_s}.
\end{align}
This equation determines the value of the ambipolar potential $\phi$ in terms of the temperatures, applied potentials, and midplane densities of each species $s$.

If there is only one ion species $c$, with mass $m_c$ and charge state $Z_c$, then everywhere in the plasma we have:
\begin{align}
	n_{eC} e^{-\psi_e / T_e} = Z_i n_{iC} e^{-\psi_c/ T_c} \Longrightarrow \frac{\psi_e}{T_e} = \frac{\psi_c}{T_c}. \label{eq:SingleSpeciesAmbipolar1}
\end{align}
Using Eq.~(\ref{eq:psiS}) to eliminate $\psi_s$ and solving for $\phi$, we find (assuming $\Psi_e = 0$):
\begin{align}
	e \phi = -\frac{T_e}{T_c + Z_c T_e} \Psi_c. \label{eq:AmbipolarPotential}
\end{align}
Using Eq.~(\ref{eq:psiS}) allows us to then determine the net potential seen by the ions:
\begin{align}
	\psi_c = \left(\frac{T_c}{T_c+ Z_c T_e} \right) \Psi_c. \label{eq:ConventionalScreen}
\end{align}
We thus see that the net effect of the ambipolar potential is to reduce the confining potential of the end plugs.

\changes{The strength of the screening can be understood by considering the limiting cases.  In the limit of $T_e \rightarrow 0$, the electrons are effectively pressureless, and thus can be manipulated by an extremely weak ambipolar potential.
This weak ambipolar potential, which enforces co-location of the electrons and ions, is barely felt by the ions relative to the very strong ponderomotive potential.
Thus the total potential on the ions is basically the same as the ponderomotive potential.}

\changes{Conversely, in the limit $T_e \rightarrow \infty$, the electron pressure dominates the ion pressure.  In this limit, the electrons must assume nearly uniform density, so the ions must then also assume nearly uniform density, implying a vanishingly small net potential on the ions.
As a result, the ponderomotive potential must be largely screened out in this case.}

\changes{What appears to be counterintuitive about the role that the electrons play is that it is precisely in the regime in which plasma shielding is most potent ($T_e \rightarrow 0$ with vanishing Debye length), that the external potential forces on the ions survives intact, like in a vacuum. The difference, of course, is that in the familiar problem of Debye screening of a probe, it is the ion charge that is prescribed, not the potential. The precise canceling of the charge without the generation of large electric fields is in fact what allows the potential to survive unaltered.}

The intermediate regimes are of course the most relevant and important for applications.  
In a hydrogenic ($Z_c = 1$) plasma with thermally equilibrated electrons, the potential is reduced by a factor of 2.

\section{Theory of the Strong-Doping Limit} \label{sec:HighDoping}

To see how things can change in a multi-species plasma, consider the limit where the ion species $c$ that we are trying to confine has negligible density, with the bulk of the plasma consisting of a different dopant species $d$.
In this case, the ambipolar potential is given by Eq.~(\ref{eq:AmbipolarPotential}) with $c \rightarrow d$.
Plugging this ambipolar potential into Eq.~(\ref{eq:psiS}) for species $c$, we find that the net potential seen by trace species $c$ is then
\begin{align}
	\psi_c = \Psi_c - \frac{T_e}{T_d + Z_d T_e} \Psi_d. \label{eq:StrongDopingPsiC1}
\end{align}
\changes{It can be seen immediately from Eq.~(\ref{eq:StrongDopingPsiC1}) that the presence of a large number of dopant ions allows the dopant to ``fill in'' for the trace ions to balance the charge imbalance; if $\Psi_d = 0$, the trace ions would see only the imposed potential (just as if the electrons were cold).  
However, if the potential $\Psi_d$ is negative, then one can actually do better, ending up with an net potential even larger than the applied ponderomotive potential.  
This motivates examining the ponderomotive potential, which can take different signs for different ion species.}

Plugging the form of the ponderomotive potential from Eq.~(\ref{eq:PonderomotiveScaling}), Eq.~(\ref{eq:StrongDopingPsiC1}) becomes:
\begin{align}
	\psi_c =  \left[ 1 - \frac{1-\bar{\Omega}_c^2}{1-\bar{\Omega}_d^2} \frac{Z_d^2}{Z_c^2} \frac{m_c}{m_d} \frac{T_e}{T_d + Z_d T_e}  \right] \Psi_c. \label{eq:EnhancementHighDoping}
\end{align}
Crucially, it is clear that the sign of the second term can be positive, if $(1-\bar{\Omega}_c^2)$ has the opposite sign to $(1-\bar{\Omega}_d^2)$--meaning that the end plug can enhance the ponderomotive potential seen by species $c$, rather than diminishing it.
\changes{Note also that high electron temperature, which reduced ion confinement in the single-species case (Eq.~(\ref{eq:ConventionalScreen})), now serves to enhance ion confinement in the doped case.}

If our goal is to form a repulsive end plug, $Z_d$ cannot be large, because to simulaneously satisfy $(1-\bar{\Omega}_d^2)<1$ while $(1-\bar{\Omega}_c^2) > 1$, we must have $Z_d/m_d > Z_c / m_c$, which generally forces $d$ to be a light element.
In practice, this means we will be looking at two options: H, and He$^{3}$.

To proceed further, it is necessary to make assumptions about how close each species is to resonance.
Of course, it is possible using Eq.~(\ref{eq:EnhancementHighDoping}) to get an arbitrarily large enhancement if put the wave frequency very close to the dopant cyclotron resonance; however, this is not very realistic or informative. 
Instead, therefore, we will assume that $(1-\bar{\Omega}_c^2)$ and $(1-\bar{\Omega}_d^2)$ have the same magnitude but opposite sign, which roughly implies that we ``avoid the resonance'' to the same degree for each species\changes{---a plausible engineering approach to design a maximally robust end plug}.
\changes{This choice leads to interprable results on the strength of the effect, but it must be noted that it is somewhat arbitrary: in a realistic device, the degree to which resonance can be approached depends on the amplitude-dependent nonlinear transition to stochastic diffusion, which breaks the ponderomotive model.\cite{Karney1978StochasticIon,Karney1979StochasticIon,Dawson1983DampingLargeAmplitude} }
Thus the wave might be targeted closer to one resonance or the other in a true optimization.

Proceeding, for deuterium confinement with a hydrogen end dopant, we have:
\begin{align}
	\psi_c =  \left[ 1 + 2 \frac{T_e}{T_d + T_e}  \right] \Psi_c.
\end{align}
If the dopant is kept cold relative to the electrons, this means there is a theoretical limit of a 3x enhancement in the ponderomotive potential due to ambipolar effects, rather than the 2x reduction found for the single-species case.
This therefore represents a 6x improvement over the single-species case.

For an He$^3$ dopant, we instead have:
\begin{align}
	\psi_c =  \left[ 1 + \frac{4}{3} \frac{T_e}{T_d/2 + T_e}  \right] \Psi_c.
\end{align}
This is a slightly weaker enhancement factor than for the hydrogen dopant.
However, note that for the He$^{3}$ case, the ponderomotive forces for both species can be put simultaneously closer to resonance, potentially resulting in a higher force at lower electric field.
To see this effect of the resonance, note that the closeness to resonance is determined by:
\begin{align}
	(1-\bar{\Omega}_c^2) = (\bar{\Omega}_d^2 - 1).
\end{align}
For $c=$D and $d=$H, the above formula (combined with the scaling from Eq.~(\ref{eq:PonderomotiveScaling})) gives $\bar{\Omega}_c = 0.632$, and thus $(1-\bar{\Omega}_c^2)  = 0.6$, which is quite far from the resonance and thus less amplified by the resonant denominator.
In contrast, For $c=$D and $d=$He$^3$, the same equation gives $\bar{\Omega}_c = 0.848$, and thus $(1-\bar{\Omega}_c^2)  = 0.28$, resulting in a 2.2x stronger resonant amplification effect.
Thus, in some scenarios, He$^3$ could be the preferred dopant.

\section{Theory of the Weak-Doping Limit} \label{sec:LowDoping}

Thus far, we have considered the case of trace confined species $c$.
However, such a theory does not say much about how much dopant is required to see a substantial effect.
Thus, we now turn our attention to the weak-doping limit.

In this limit, we can take the central-cell dopant density $n_{dC}$ to be 0, while keeping a finite density for the dopant density $n_{dE}$ in the end cell.
We can then solve a modified equation for the ambipolar potential:
\begin{align}
	n_{eC} e^{-\psi_e / T_e} = Z_c n_{cC} e^{-\psi_c / T_c} + Z_d n_{dE} . \label{eq:LowDopantStart}
\end{align}
This model will be valid as long as 
\begin{align}
	\frac{Z_d}{Z_c} \frac{n_{dE} e^{\psi_d/T_d}}{n_{cC}} \ll 1. \label{eq:LowDopantValidity}
\end{align}

We can recast this equation in terms of the modification to our single-species solution, i.e. take $\phi = \phi_0 + \phi_1$, with $\phi_0$ given by Eq.~(\ref{eq:AmbipolarPotential}). 
Then, Eq.~(\ref{eq:LowDopantStart}) becomes:
\begin{align}
	e \phi_1 \left( \frac{1}{T_e} + \frac{Z_c}{T_c} \right)&= \log \left(1 + \frac{Z_d}{Z_c} \frac{n_{dE}}{n_{cC} }e^{e\phi_{0}/T_e + Z_c e\phi_1/T_c}\right). \label{eq:phi1LowDoping}
\end{align}

\begin{figure}
	\centering
	\includegraphics[width=\linewidth]{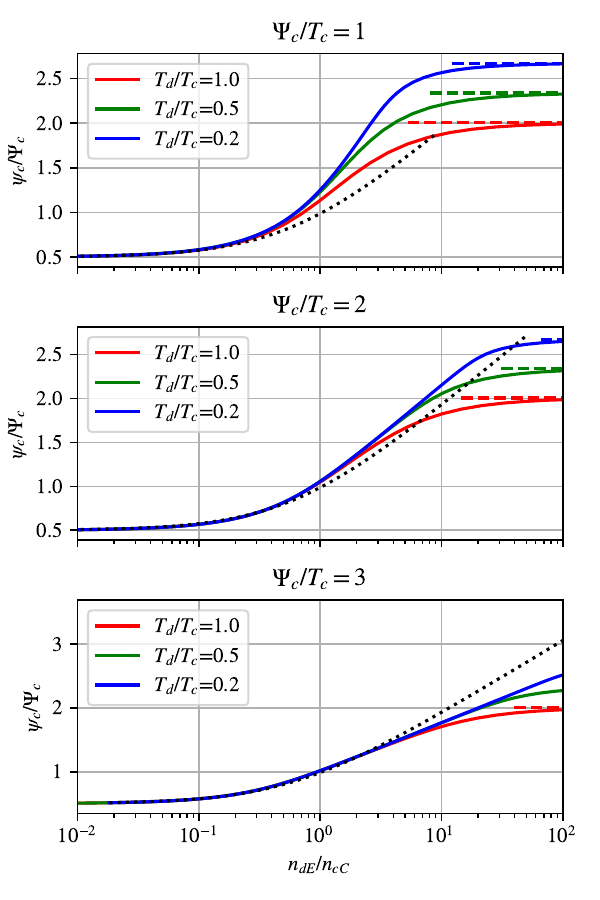}
	\caption{Enhancement in total potential experienced by a deuterium plasma due to hydrogen doping of the ponderomotive end cells, for several values of the ponderomotive potential $\Psi_c$ and of the dopant temperature $T_d$. It is assumed here that electrons and deuterium are thermally equilibrated. 
		The x-axis shows the degree of doping, i.e. the ratio of the end cell density of the hydrogen dopant relative to the central cell density of deuterium.
		The y-axis shows the ratio of the total effective potential $\psi_c$ seen by the deuterium (ponderomotive + ambipolar) to the raw applied ponderomotive potential $\Psi_c$.
		The solid line corresponds to the numerical solution (Sec.~\ref{sec:Numerics}), the dashed line to the strong-doping limit (Sec.~\ref{sec:HighDoping}), and the dotted line to the weak-doping limit theory (Sec.~\ref{sec:LowDoping}).
		The theories agree well in the appropriate limits, and it is apparent that doping can lead to significant enhancement of the ponderomotive potential.
		Futhermore, is clear that a cold dopant allows for somewhat better performance.}
	\label{fig:HDopedDPlug_psiC}
\end{figure}

\begin{figure}
	\centering
	\includegraphics[width=\linewidth]{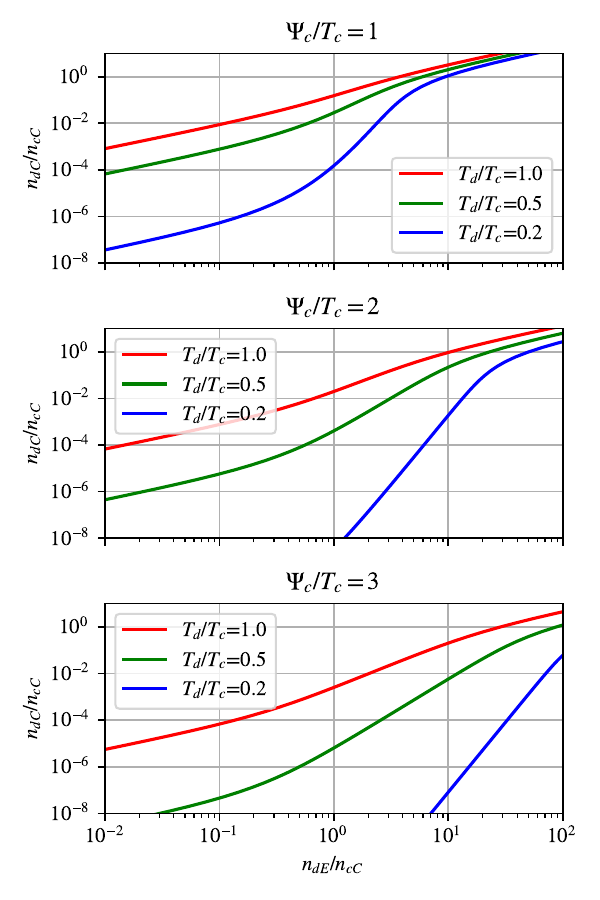}
	\caption{Ratio of central cell dopant density $n_{dC}$ to central confined species density $n_{cC}$, for the same cases as shown in Fig.~\ref{fig:HDopedDPlug_psiC}.
		Comparing the two figures, we see that it is possible to get a high degree of enhancement of the ponderomotive potential even with negligible poisoning of the central cell plasma by the dopant, especially if the dopant is kept relatively cold.}
	\label{fig:HDopedDPlug_nd0}
\end{figure}

At low values of $n_{dE}$, we can ignore the $\phi_1$ factor in the exponential on the RHS, giving a conventional linearized solution.
However, the equation is more useful in revealing when the enhancement factor becomes large.
In the limit of large $\phi_0$, i.e. a strong ponderomotive potential, if we let $Z_d n_{dE}/Z_c n_{cC} \rightarrow 1$, then we find:
\begin{align}
	\phi_1 = - \phi_0.
\end{align}
In other words when the dopant charge density in the end cell reaches the density of the confined species in the central cell, the ambipolar potential disappears.
Intuitively, this makes sense, as this is the point at which the electrons see a ``flat'' charge profile as they traverse both cells.
At dopant densities higher than this, the ambipolar potential is determined more by the dopant than the confined species, and provides a confining effect on the plasma; i.e. when:
\begin{align}
	\frac{Z_d}{Z_c} \frac{n_{dE}}{n_{cC}} \gtrsim 1; \label{eq:SignificantEnhancement}
\end{align}

%
%

Importantly, substantial enhancement can be achieved even when the dopant density in the central cell is negligible, since conditions~(\ref{eq:LowDopantValidity}) and (\ref{eq:SignificantEnhancement}) can be simulataneously satisfied due to the exponential factor.
Notably, it is easier to simultaneously satisfy both conditions at low dopant temperature.

\section{Numerical Validation} \label{sec:Numerics}

Leaving behind approximations, the full equation to be solved in a two-ion-species plasma is:
\begin{align}
	n_{eC} e^{-\psi_e / T_e} = Z_c n_{cC} e^{-\psi_c / T_c} + Z_d n_{dC} e^{-\psi_d / T_d},
\end{align}
with 
\begin{align}
	\psi_e &= -e \phi\\
	\psi_c &= Z_c e \phi + \Psi_c \\
	\psi_d &= Z_d e \phi + \Psi_d,
\end{align}
and where the ponderomotive potentials of $\Psi_c$ and $\Psi_d$ are related by:
\begin{align}
	\Psi_d &=  \frac{Z_d^2}{Z_c^2} \frac{m_c}{m_d} \frac{1 -\bar{\Omega}_c^2}{1-\bar{\Omega}_d^2} \Psi_c.
\end{align}
As before, we assume $(1-\bar{\Omega}_c^2) = -(1-\bar{\Omega}_d^2)$.
Thus, to define an equilibrium, we set $n_{sC}$, $T_s$, and $\Psi_c$.
$\phi$ is then determined by the ambipolar conditions.
At the end, we can check what the net confining potential $\psi_c$ is in the off-midplane region, as well as the density of the dopant in the end cell.

In Fig.~\ref{fig:HDopedDPlug_psiC}, we show the results of the numerics for confined deuterium with a hydrogen dopant, for several values of the ponderomotive potential and dopant temperature, assuming that the electron and deuterium temperatures are equal, i.e. $T_c = T_e$.
Although we scan over the dopant density $n_{dC}$ at the midplane, we plot results in terms of the dopant density $n_{dE}$ at the end cell, relative to the confined species density, since this was shown to be the critical parameter in Sec.~\ref{sec:LowDoping}.
It is clear that the weak doping theory accurately describes the results until $n_{dE}/n_{cC} \sim 1$.
Furthermore, the strong-doping theory accurately predicts the maximum possible achievable ponderomotive enhancement.

It is very clear from the above analysis and numerical results that there is an advantage to having a low dopant temperature.
Keeping the dopant cold has two beneficial effects. 
First, low dopant temperature increases the maximum enhancement factor in the strong-doping limit, as seen in the high-$n_{dE}$ limit of Fig.~\ref{fig:HDopedDPlug_nd0}.
Second, and perhaps even more significantly, it decreases the density of the dopant in the central cell for a given end cell dopant density (Fig.~\ref{fig:HDopedDPlug_nd0}), so that substantial confinement can be achieved with minimal dopant penetration into the confined plasma.


\section{Discussion and Conclusion}

In this paper, we have shown that there is a large possible upside potential to using lightweight dopants in ponderomotive end plugs, which can eliminate or even reverse the typical ambipolar screening of an undoped plug.
While promising, it is important to note that there are many assumptions in the model employed here, which could affect the application of the results.

First, the enhancement factor was calculated assuming that the two ion species were placed equally close to the cyclotron resonance (as measured by the strength of the resonant denominator in the ponderomotive potential).
However, in an optimized configuration, it might make sense to place one species or the other closer to resonance.
For instance, since applying the potential to the hydrogen dopant often results in a larger net potential, one might choose to place the wave frequency much closer to the hydrogen cyclotron frequency.
How close one can get to this frequency---and thus how large one can make the resulting potential for a given electric field amplitude---depends on the criterion for the onset of stochastic diffusion, which can depend on the wave amplitude\cite{Karney1978StochasticIon,Karney1979StochasticIon} or the collision frequency.\cite{Swanson1989PlasmaWaves,Stix1992WavesPlasmas}

Second, our analysis assumed that the wave was transversely polarized.
However, it is more typical for waves propagating into a dense plasma, particularly near to the ion cyclotron frequency, to be elliptically polarized.\cite{Stix1992WavesPlasmas}
This change in polarization can dramatically change the ponderomotive potentials.\cite{Dodin2005ApproximateIntegrals,Rubin2024FlowingPlasma,Rubin2025PonderomotiveBarriers}
Thus, in the detailed design of a ponderomotive end plug for a dense plasma, the application of the RF wave must be self-consistently taken into account.
\changes{Some implications of considering these more general polarizations, including for static end plugs, are dicussed in Appendix~\ref{app:GeneralPolarization}.}

Importantly, we have assumed here that the plasma is being operated in a mode where the ponderomotive potential is applied at the ends of the device, creating a repulsive potential for the confined species.
However, if one is trying to confine a light species such as hydrogen, it is also in theory possible to place an attractive potential in the central cell, and a heavy dopant at the edges, in effect reversing the roles of central cell and end plug in the analysis in the paper.
In this way, one might be able to access the advantage of end doping even if one is trying to confine a light species.
\changes{In general, one can even apply a combination of an attractive ponderomotive potential in central cell and a repulsive ponderomotive potential in the end cell to boost confinement further,\cite{Rubin2025PonderomotiveBarriers} though it should be noted that for transverse waves, Eq.~\ref{eq:PonderomotiveScaling} implies that it is only possible to enhance one of these potentials with the use of a single dopant.}

\changes{Finally, it should also be noted that all of the calculations here pertain to the one-dimensional distribution of particles along a given magnetic field line. 
However, any redistribution of the potential along a field line can also modify the distribution of the electrical potential across field lines. 
This can drive or (more generally) modify plasma rotation. 
Hence, the same mechanisms that are employed in this paper to enhance confinement of a species (e.g., by imposing ponderomotive potentials and varying the relative temperatures of different species) could also be used to control plasma rotation. 
This possibility may be especially interesting in contexts where the shear profile is important. \cite{Piterskii1995StabilizationDriftcone,Hassam1999VelocityShear,Huang2001VelocityShear,Cho2005ObservationEffects,Maggs2007TransitionBohm,Beklemishev2010VortexConfinement}
In a rapidly rotating system (for instance, one in which the rotation speeds are supersonic), the cross-field voltage drops are often large compared with the parallel voltage drops associated with the ambipolar fields considered here. 
However, there are parameter regimes in which the parallel electric fields are important. \cite{Kolmes2024MassiveLonglived}
In these regimes, the ideas described in this paper could provide alternative, non-resonant techniques for wave-based rotation control (as opposed to resonant mechanisms like alpha-channeling \cite{Fisch1992InteractionEnergetic,Fetterman2008AlphaChanneling,Ochs2021WavedrivenTorques,Ochs2022MomentumConservation,Ochs2024WhenWaves}). 
The full extent to which these effects provide practical ways with which to control rotation goes beyond the scope of this paper. }

Clearly there are many open questions still to explore in unlocking the utility of ponderomotive end plugs.
Nevertheless, it is apparent that end doping has the potential to dramatically improve the confinement potential of this technology.

\section*{Acknowledgements}

This work was supported by Princeton University.

\section*{Data Availability}

Data sharing is not applicable to this article as no new data were created or analyzed in this study.

\appendix

\section{Implications of Ambipolar Fields for More General Scenarios} \label{app:GeneralPolarization}

Most of this paper focuses on the ponderomotive potentials associated with transverse polarization. 
It also focuses, to a large extent, on the case that is at least moderately close to resonance. However, there are a number of nontrivial effects that arise when we consider other limits. 

For Cartesian $(x,y,z)$ coordinates and a $z$-directed magnetic field, we can decompose an applied RF electric field into parallel and circularly-polarized components as follows: 
\begin{align}
	E_{||} &= E_z \\
	E_+ &= \frac{1}{\sqrt{2}} \big( E_x - i E_y \big) \\
	E_- &= \frac{1}{\sqrt{2}} \big( E_x + i E_y \big) . 
\end{align}
In terms of these field components, the resulting ponderomotive force can be written as\cite{Motz1967RadioFrequencyConfinement}
\begin{align}
	\Psi_s = \frac{Z_s^2 e^2}{4 m_s \omega} \bigg( \frac{|E_+|^2}{1+\bar \Omega_s} + \frac{|E_-|^2}{1-\bar \Omega_s} + |E_{||}|^2 \bigg). 
\end{align}
Here $Z_s e$ is the charge of species $s$, $e$ is the elementary charge, $\omega$ is the frequency of the applied field, and $\bar \Omega_s$ still denotes the ratio of the gyrofrequency of species $s$ and $\omega$. 
For the transverse polarization considered in most of this paper, $|E_+| = |E_-| = |E|/\sqrt{2}$ and $|E_{||}|=0$, in which case 
\begin{gather}
	\Psi_{s,\text{transverse}} = \frac{Z_s^2 e^2}{4 m_s \omega} \frac{|E|^2}{1-\bar{\Omega}_s^2} \, . 
\end{gather}
This recovers the scaling from Eq.~(1). 

For more general polarizations, the most immediate difference is the dependence on $\bar \Omega_s$. 
For example, this paper discusses interesting effects that can arise when different ion species see $\Psi_s$ with different signs. 
If $\omega > 0$ and $\Omega_s > 0$, the sign of $\Psi_s$ can be different for different ion species only in the part driven by the $E_-$ component. 
Moreover, this paper discusses the enhancement of $\Psi_s$ that happens when $\bar \Omega_s^2$ approaches 1. 
The scaling of this enhancement (not to mention its dependence on the sign of $\bar \Omega_s$) clearly depends on the choice of polarization, and disappears altogether for the parallel polarization. 

It is also interesting to consider the behavior of the ponderomotive potential in cases where $\bar \Omega_s \gg 1$ and $E_{||} \rightarrow 0$. In this limit, 
\begin{gather}
	\Psi_s = \frac{Z_s e c}{4 B} \bigg[ |E_+|^2 - |E_-|^2 + \mathcal{O} \bigg( \frac{|E|^2}{\bar \Omega_s} \bigg) \bigg] . 
\end{gather}
In this case, $\Psi_s$ vanishes at leading order in $\bar{\Omega}_s^{-1}$ if $|E_+| = |E_-|$. 
But perhaps more interestingly, $\Psi_s \propto Z_s$ and -- again, to leading order in $\mathcal{O}(\bar \Omega_s^{-1})$ -- it has no mass dependence. 
Then following the argument from Section~II, the self-consistent ambipolar fields exactly cancel the leading-order ponderomotive potentials $\Psi_s$, so that the net potentials $\psi_s$ vanish. 
Intuitively, in this limit $\Psi_s$ has the same scaling as an electrical potential.
This means that the ponderomotive potentials from low-frequency transverse waves have \textit{no net effect} on the overall plasma confinement once an ambipolarity condition is imposed. 

The high-$\bar{\Omega}_s$ limit happens to be particularly relevant for static-field ponderomotive endplugs. 
In static-field endplugs, the ponderomotive effect could be driven by the flow of the plasma over a static field perturbation\cite{Rubin2023MagnetostaticPonderomotive, Ochs2023CriticalRole, Kolmes2024CoriolisForces, Rubin2024FlowingPlasma, Rubin2025PonderomotiveBarriers} (e.g., a rotating plasma with a field perturbation that has some azimuthal mode number), resulting in an effective RF field in the reference frame co-moving with the plasma. 
For these configurations, the penetration of field perturbations into the interior of the plasma is a major constraint, and in some cases, these propagation-related issues for transverse polarizations favor large $\bar{\Omega}_s$.\cite{Rubin2024FlowingPlasma} 
The results in this paper suggest that this limit may not be useful for endplugging quasineutral plasmas. 
This does not preclude static-field endplugs generally, but it does highlight the importance of self-consistent ambipolar fields for the evaluation of different endplug configurations.

%

\clearpage
\newpage


\end{document}